# Selective Pumping of Localized States in a Disordered Active Medium

Bhupesh Kumar[1], Mélanie Lebental[2] and Patrick Sebbah[1,3]

[1]Department of Physics, The Jack and Pearl Resnick Institute for Advanced Technology, Bar-Ilan University, Ramat-Gan, 5290002 Israel

[2]Laboratoire de Photonique Quantique et Moléculaire, ENS Cachan, CentraleSupélec, CNRS, Université Paris-Saclay, 94235 Cachan, France

[3]Institut Langevin, ESPCI ParisTech CNRS UMR7587, 1 rue Jussieu, 75238 Paris cedex 05, France

**Abstract:**

**Light scattering and localization in strongly scattering disordered systems is governed by the nature of the underlying eigenmodes, specially their spatial extension within the system. One of the main challenges in studying experimentally Anderson-like localized states resides in the difficulty to excite these states independently and observe them individually anywhere in the sample. The modes' mutual interaction and their coupling to the sample boundaries makes it extremely challenging to isolate them spectrally and image them alone. In the presence of gain, random lasing occurs above threshold and lasing modes are excited. It was shown recently that shaping the intensity profile of the optical pump is a very effective way to control and force random lasing in singlemode regime at any desired emission wavelength. Active random media therefore offer a unique platform to select modes individually. By using pump shaping technique in a strongly scattering random laser, we successfully select localized lasing modes individually. Direct imaging of the light distribution within the random laser confirms the confined nature of the modes and enables us to elucidate long-standing questions on the role of non-Hermiticity in active scattering media and the influence of nonlinearities on Anderson wave localization.**

Real eigenvalues and orthonormal eigenfunctions are hallmarks of Hermitian systems. But when a wave system starts to couple to its surroundings or when a non-uniform distribution of gain or loss is introduced, the modes become complex-valued and their eigenfunctions are no longer orthogonal. Being ubiquitous in nature, Non-Hermitian potentials have attracted considerable attention recently. They form a new paradigm where new degrees of freedom and unprecedent control over wave propagation and dynamics are offered, leading to unexpected new features [1,2,3]. Disordered media are an interesting example of non-Hermitian system, where the amount of non-Hermiticity can be tuned by varying the scattering strength of the medium. If scattering is sufficiently strong, transport is suppressed and modes are exponentially localized, away from the sample limits. This is the celebrated regime of Anderson localization [4], which has become an important frame for the understanding of transport in mesoscopic systems and classical-wave multiple scattering. As disorder strength is lessened, modes spatially extend and couple to their surroundings, while spatial and spectral overlap between modes become substantial.

Introducing gain may reveal further the extreme sensitivity of the system to mode complexness [5] and modal non-orthogonality [6]. The observation of sharp lasing peaks in semiconductor powders has triggered the interest for random lasing, raising the question of the origin of the laser oscillation and the nature of the lasing modes in this otherwise mirrorless laser [7]. In the regime of Anderson localization, the lasing modes are predicted to build up on the Anderson-localized modes of the passive system [8]. It was further realized that random lasing can also occur in diffusive samples [9]. In that case, complexness leads to random lasing modes which can be fundamentally different from the modes of the passive system [10,11,12]. Because of the complexity of the problem, theoretical and numerical studies have mostly focused on near-threshold operation, where gain saturation and mode competition for gain can be ignored. A full ab-initio method is otherwise required to find the lasing modes [13]. To date, no experimental investigations have supported these predictions.

Lasing in the regime of Anderson localization is also difficult to achieve and to demonstrate experimentally. It was first investigated in a 1D stack of microscope cover slides with active dye [14], where a low-threshold laser emission was assumed to occur on long-lived modes overlapping spatially with the gain region, without direct evidence of the nature of the modes. [15,16,17] The regime of Anderson localization of light has been proposed to obtain stable multimode random lasing [18], which was demonstrated in a photonic crystal membrane embedded with quantum wells [19]. But to date no experiment has been able to identified unequivocally localized lasing modes and to compare them with the modes of the passive system [8]. The experimental challenge resides in the highly multimode nature of random lasers. Spectral and spatial modal overlap makes it impossible to isolate and image the modes individually. Nonlinear modal interaction and strong gain-induced nonlinearities, saturation and cross-saturation makes it even more difficult to infer anything from the measurements. Only in the particular situation of very strong scattering, as it is the case for transverse localization, does the localized modes have been observed individually [20].

Random lasers are extremely sensitive to local or non-uniform pumping [12,21]. It was shown recently that, by shaping the gain distribution, cross-saturation can be controlled and singlemode operation achieved at any desired selected emission wavelength [22,23]. This new method to control laser emission has also been successfully applied to microdisk lasers [24] and asymmetric resonant cavities [25]. It has been proposed to control spectral characteristics [26], power efficiency [27], emission directionality [28,24], laser threshold [29] and modal interactions [30,31,32] in various types of 2D lasers and to single out whispering-gallery modes in micro-resonators [25]. In this work, we propose to use this method to reveal the disordered-induced localized nature of the lasing modes in a strongly scattering disordered medium in the presence of gain. Here, a one-dimensional solid-state random laser is designed, where non-Hermiticity can be tuned by varying the degree of scattering as well as by shaping the gain medium by nonuniform pumping. By applying the pump shaping method to suppress modal interaction and select a particular lasing mode and by imaging directly the emission along the active random structure, we find that all lasing modes are spatial confined. We demonstrate experimentally that these lasing modes built up on the passive modes, and therefore identify the the disordered-induced localized modes of the systems. We check their sensitivity to the sample boundaries [Thouless] by measuring the increase of their lasing threshold near the sample edge and found the optimally-outcoupled lasing mode, which has the largest slope efficiency. Their robustness to non-uniform pumping and nonlinear gain is investigated and we proove experimentally that localized stated are insensitive to nonlinear optical gain and nonlinear interactions.

To conduct our investigation, we have designed a solid-state organic random laser (Fig.1a) by spin-coating a 600 nm-thin layer of doped-polymer (DCM-doped PMMA) on a quartz plate [see Methods]. Once polymerized, this layer is structured by engraving parallel randomly-spaced grooves using electron-beam lithography. Spontaneously emitted light guided within the polymer film is scattered and amplified as it propagates along the structure, when optically pumped transversally at 532 nm. This solid-state random laser, dubbed "the plastic random laser", benefits from laser stability, sample longevity and most importantly for our study, relatively high index contrast between air and polymer ($\Delta n = 0.65$)[33], which is favorable to reach the regime of Anderson localization within the limit of the sample. Above all, the structuration of the sample offers a bonus: out-of-plane scattering allows direct imaging of the near-field intensity profile within the laser, as seen in Fig. 1b. Controlled disorder is introduced by varying the groove position around a periodic arrangement. Finally, the degree of scattering and, consequently the localization length, can be adjusted by varying the depth of the grooves. In the present study, the scattering system is composed of 125 identical parallel grooves, randomly positioned around an 8 µm-pitch periodic lattice with standard deviation 3.51 µm (Fig. 1c). Each groove is 600 nm-deep, 50 µm-long, and 350 nm-wide (see more details about the geometry in SM). The total length of the sample is $L$=1000 µm. With these parameters, the system is well approximated by a one-dimensional layered system and the localization length is found numerically to be $\xi \sim 65$ µm [see SM].

The schematic of the experimental set up shown in Fig. 2f. The 532 nm beam of a frequency-doubled Nd:YAG picosecond laser (Ekslpa pl2231-50) is formed into a 50 µm x 500 µm stripe to pump uniformly at normal incidence the scattering structure. Emitted light along the waveguide is collected at the sample edge by a microscope objective and fiber-coupled to a high-resolution spectrometer (Horiba IHR500 equipped with a E2V Synapse CCD camera). At the same time, the surface of the sample is imaged through a microscope objective onto a CCD camera (Andor Zyla 4.2+). Lasing action is observed for pump energy above 18 nJ, when sharp discrete peaks appear in the emission spectrum (see Fig. 2c). As it is often the case with random lasers, the laser emission is multimode above threshold (Fig. 2d). The emission spectrum is stable at all lasing wavelengths and independent of the pump fluence. This is seen in Fig. 2d, where the emission wavelength of each lasing mode is found to be constant at all values of the pump energy. This is an important observation, which means that lasing modes in our system are extremely robust to gain and other type of optical nonlinearity.

The image of the sample surface observed under the microscope objective shows intense out-of-plane scattered light at the position of the air grooves when the system is pumped above threshold (Fig. 2a). This is in contrast to below-threshold pumping, where weak uniformly-distributed intensity is detected, which corresponds to amplified spontaneous emission (ASE) (Fig. 2b). By monitoring independently and simultaneously the spatial distribution of both laser intensity and spontaneous emission, we demonstrate here the clamping of spontaneous emission. This feature, which has never been shown for random lasers, is a hallmark of lasing and occurs when stimulated emission becomes the preferred transition mechanism. Clamping of spontaneous emission is shown in Fig. 2e where the growth of spontaneous emission is seen to be drastically curtailed as the threshold is reached and laser oscillations take over.

Because all grooves are identical at optical scale, the sampled-intensity profile of the optical field within the random laser is obtained from the image (Fig. 2a), after the spontaneous-emission contribution is subtracted (See SM). A typical profile of multimode random lasing intensity along the uniformly-pumped sample is shown in Fig. 3a. Several modes contribute to this profile, preventing any investigation of their spatial distribution individually. To observe them separately, we need to operate the random laser in singlemode operation at the lasing wavelength of the mode we want to investigate (inset of Fig. 3a). This is achieved by shaping the intensity profile of the pump following the method proposed in [23] in order to excite selectively this particular mode. After reflection on a computer-driven spatial light modulator (SLM), the pump beam is modulated and projected on the back-side of the sample, as sketched in fig.2f (see Methods for more details). Because the pump profile that selects a particular mode is not known a priori, an iterative method based on optimization algorithm is used (see Methods) to find the optimal pump intensity profile that selects a particular mode in the multimode lasing spectrum. The method is illustrated in Fig. 3b with the selection of a single lasing mode at 599.80 nm, a mode which is not the first mode to lase under uniform pumping. After 200 iterations, a non-uniform pump profile is found (Fig. 3b), which optimally select the targeted wavelength and reject all other undesired lasing modes below threshold (Fig. 3c). Remarkably, pure singlemode operation is achieved. This in contrast to [23] where sidelobe rejection was not total, definitely because scattering is strong in our plastic laser, as anticipated in [23]. We are now in a favorable position to directly investigate a particular random lasing mode. The spatial distribution of the laser-field intensity corresponding to the selected mode at 599.80 nm is measured and shown in Fig. 3b. The lasing mode is found to be spatially confined away from the sample limits. Multiple scattering and disordered-induced localization is shown here to provide with the necessary optical feedback to reach threshold, as an optical cavity would do in a conventional laser. The same operation is reproduced to select all other lasing modes from the multimode emission spectrum of Fig. 2c and to investigate their spatial profile. We confirm that all lasing modes are spatially localized. Figure 4 shows six examples of lasing modes spatially localized in different regions of the sample, together with the corresponding optimized pump profile. A systematic study of the spatial extension of the lasing modes is presented in SM, which show strong fluctuation of the localization length around the predicted value found numerically.

Observation of localized optical modes has been previously reported for regime of very strong scattering, where modes were strongly confined and do not overlap with each other [19,34]. In our system, localization length is a fraction of the sample size and therefore modal overlap is significant, as seen in Fig. 4, suggesting mode competition for gain and nonlinear mode coupling. Nevertheless, the pump shaping method reveals itself extremely efficient. Indeed, the optimization algorithm does not trivially converge to a narrow pump profile near the center of the mode, which would been sufficient to spatially select strongly confined modes [8,22]. Here, the distribution of the pump intensity extends beyond the center of the mode. Optimal selection of the lasing mode is therefore a complex process, where suppression of gain cross-saturation and mode coupling is at work in the selection operation. This is confirmed in Fig. 3d by comparing the laser characteristic of the mode for uniform and optimized pump profile, respectively. When the system is pumped uniformly (multimode regime), the targeted mode emission @ 599.78 nm is impeded as cross-saturation effects are at play between different lasing modes. Strikingly, the threshold of the lasing mode after optimization is reduced by a factor 1.5, while the slope efficiency of its laser characteristic is increased by a factor six when compared to the same mode lasing in multimode regime under uniform pumping. This means that the lightwave couples out of the random laser more efficiently when the mode is optimized. One can conclude that the selective method employed here (a) "focuses" the pump to reduce threshold, and (b) unleashes the selected mode from competing for gain with other modes, which increases its output coupling [27]. Interestingly enough, the performance of the laser has been improved by reducing the gain surface, a rather counterintuitive effect.

To investigate more thoroughly this point, the threshold and the slope efficiency have been systematically measured for 15 individual lasing modes and correlated with their position within the sample (here defined as the position of its maximum) (Fig. 5). One expects that near the sample boundaries, the threshold grows as leakage out from the scattering medium increases the loss of the mode. This is indeed what we observe for mode near the edges of the sample (red plot in Fig. 5). In turn, the measure of the threshold provides a direct evaluation of the sensitivity of the localized modes to the boundary of the system [35]. The slope efficiency presented in Fig. 5 (blue plot) shows another interesting feature: It exhibits two maxima (at ~250 µm and 700 µm), almost symmetric around the center of the sample (500 µm). Maximum slope efficiency defines the *optimally-outcoupled* lasing modes for this particular disordered system. It is worth noting that this happen neither for lasing modes near the center of the sample, nor for modes near the sample edges. Indeed, the spatial position of the *optimally-outcoupled* mode is a trade-off between low loss (mode localized at the sample center away from the sample edges) and strong output coupling (mode localized at the sample edges). This demonstrates how pump-shaping mode selection can turn on the most efficient lasing modes of a particular random laser, supporting earlier theoretical prediction [27].

Until now, we identified the lasing modes of our system individually and showed that they are spatially localized. However, unless we prove that the lasing modes we observe correspond to the modes of the passive system, we cannot infer anything on the nature of the latter. Indeed, the disordered cavities that provide feedback in random lasers are in general widely open. Their modes are determined by a non-Hermitian operator and do not form an orthogonal basis. The lasing modes and the modes of the passive cavity are therefore likely to be different, the latter being sensitive to the openness of the system. This question has been explored numerically in several theoretical and numerical works and reviewed in [10]. In the limit of extended modes in weakly scattering media, it was shown that the lasing modes can be radically different from the modes of the passive system, calling on an for an ab-initio laser theory [36]. This is in contrast to high-finesse optical cavity usually considered in conventional laser theory [37], where lasing modes are identified with the modes of the passive cavity selected by the gain, aside from a correction on their emission frequency (frequency pulling). A similar simplification should be possible in strongly scattering media where modes are localized. Because they are spatially confined away from the boundaries, localized states are insensitive to the openness of the system. One can expect therefore that the localized modes will provide the lasing modes of the active medium, even though the system is open. This was predicted numerically in [8,38]. Here, we demonstrate it for the first time experimentally.

Because the modes of the passive system are short-live and therefore not accessible to the experimental measure, a one-to-one comparison with the lasing modes is not possible. Here, we resort to an indirect method based on local pumping. If the optimally-pumped lasing mode is not a single gain-selected mode of the corresponding passive system but a particular combination of constant flux(CF)-states, then restricting the pump to a small area will certainly result in the excitation of a different combination of CF-states. The effect of partial pumping on random lasers has been reviewed in[39]. Here we show that a local pumping locally excites a lasing mode at the same frequency and with the same shape, simply like the gain medium (e.g. the Nd:YAG rod) of a conventional laser may occupy a small fraction of the total volume of the optical cavity and still select and stimulate a mode extended in the cavity.

First, we select a lasing mode of the random laser at $\lambda = 600.20$ nm, localized near the middle of the sample. Its intensity distribution is shown in Fig. 6a & 6b, as well as the associated optimized pump profile. Next, we pump the same system but this time locally, with a 40 $\mu$m-long pump stripe, near the center (Fig. 6c & 6d). At pump energy 75 nJ, a lasing mode is excited, exactly at the same wavelength ($\lambda = 600.20$ nm) as the optimized mode. Remarkably enough, its spatial profile resembles the profile of the optimally-pumped lasing mode. The correlation coefficient, which measures the similarity of the two profiles, is $r = 82\%$. The difference is attributed to loss, which significantly reduces the mode amplitude beyond the pumping area. To support this interpretation, we show that, by adding a small uniform pump over the whole sample, we can undo the loss. The profile of the selected mode is now almost perfectly recovered (Fig. 6e & 6f), with a correlation coefficient $r = 96\%$. That local pumping is sufficient to excite the optimally-pumped lasing mode prooves that the mode of the passive system has been selected by the gain, in agreement with earlier theoretical prediction [8]. This is under the assumption that the mode is far from the sample boundaries so that the system can be considered as Hermitian.

Non-Hermiticity and its impact on the nature of the mode selected by the gain can be probed at the edges of the sample. Unfortunately, it is difficult to excite selectively such leaky modes as their high threshold is out of reach in the experiment. Nevertheless, we will show a measurable effect for a mode close to but not at the sample boundary. The mode we select ($\lambda = 605.21$ nm) is centered around $x = 750$ $\mu$m, near the right edge of the sample ($x = 1000$ $\mu$m), as shown in Fig. 7a & 7b. The same system is subsequently pumped by a 42 $\mu$m-long local pump stripe near $x = 750$ $\mu$m. A new lasing mode is excited at the same wavelength, which this time poorly resembles the original selected mode. The correlation between the two profiles yields a coefficient $r = 56\%$. By undoing the loss with an additional uniform pump, the correlation between the 2 modes improves to reach $r = 90\%$. Yet, this residual discrepancy is a signature of non-Hermiticity. The solutions of the laser equations can be significantly different than the quasimodes of the passive system because modes close to the edge are short-lived, complex-valued, and not anymore orthogonal to each other.

In conclusion of this section, the iterative method we use to select the random lasing modes allows a direct investigation of the modes of the passive system, as long as they are selected away from the boundaries. The localized nature of these modes is therefore confirmed and their spatial intensity profile can be investigated further.

With the ability to control it in single or multimode mode regime, the localized random laser becomes a paradigm to explore a very controversial question, namely the impact of nonlinearities on Anderson localization Indeed, it brings naturally together localized modes and nonlinear effects which have never been considered, such as nonlinear gain, gain saturation and mode competition.

In a first experiment, the effect of nonlinear gain and gain saturation is tested on a single localized mode by increasing the pump energy and by monitoring the mode profile. We first select a mode ($\lambda = 600.20$ nm) localized near the center of the sample. We then pump this mode near its center with a small 10 $\mu$m-long uniform pump stripe. Here we prefer a focused local pump over an optimized pump profile, in order to increase the energy density and make sure gain saturation can be reached in this experiment without sample damage. We know from our previous discussion (Fig. 6), that such a small pump, although not optimized, excites quite accurately the targeted lasing mode. The pump energy is increased stepwise from threshold (14 nJ) to saturation (147 nJ) and beyond (200.3 nJ), and the lasing mode profile is recorded, as well as the laser emission spectrum (Fig. 7). We observe that the mode profile is perfectly preserved, while its intensity has increased by more than three orders of magnitude and dye saturation has been reached above 147 nJ, as shown in Fig. 8. A quantitative comparison between the initial profile measured at low pump energy and the intensity profile at saturation gives a correlation coefficient of 97%. Lasing wavelength, as well, does not shift and single mode regime is perfectly maintained. This is clear and direct evidence that localized modes are insensitive to nonlinear gain and robust to gain saturation. Interestingly, the random laser remains purely singlemode at any pump energy, once the mode has been selected. This has been tested and confirmed for all modes of the system.

In a second experiment, we monitor the profile of a localized mode in nonlinear interaction with another mode. To do so, we carefully identify two lasing modes (mode 1 at $\lambda_1 = 608.0$ nm and mode 2 at $\lambda_2 = 602.72$ nm) with some degree of spatial overlap ($r = 33\%$), together with the corresponding optimized pump profiles, $P_1$ and $P_2$, which select each of them. First, mode 1 is excited just above threshold by applying pump $P_1$ (48 nJ). Next, we apply the second pump profile $P_2$ and, while the pump energy $P_2$ is progressively increased, we monitor both the emission spectrum and the laser-field intensity distribution across the sample. The evolution of the emission spectrum (shown in Fig. 10) dramatically illustrates the complex nonlinear competition and the mutual cross saturation effect at stake between the 2 modes. When pump $P_2$ is turn on, mode 1 takes advantage of this new resource to grow. This is made possible because the pumping profiles $P_1$ and $P_2$ have ben chosen to overlap spatially. This happens at the expense of mode 2, which is forced to remain below threshold, at a pump energy larger than its threshold when pumped alone at 55 nJ. When the energy of $P_2$ reaches 80 nJ, mode 2 starts to lase and to compete for gain. It saturates the gain provided by $P_2$ and forces mode 1 to decrease. This is a surprising demonstration of interaction-induced mode switching [40]. At a pump energy of 100 nJ, mode 1 takes control again and saturates mode 2. This no-win scenario of gain competition reproduces itself at 125 nJ, and 140 nJ. One may assume that such a strong gain competition would affect the spatial profile of the modes and would influence localization. Actually, this is not the case at all. The intensity profile resulting from the simultaneous pumping at $P_1$ and $P_2$ at any value of $P_2$ is compared to the profiles of the modes selected individually as shown in Fig. 11. We show that it is precisely the sum of the profiles of each of the modes. The mode profile is therefore unaffected by cross-modal competition.

In conclusion, we have explored random lasing in the strongly scattering regime, and demonstrated disordered-induced localization, by combining selective pump shaping and imaging of individual lasing modes in a newly-designed plastic random laser. The impact of the openness of the system and the non-uniform pumping has been explored. This led us to the conclusion that the modes of the passive systems provide with the necessary feedback for laser oscillations and are identical to the corresponding lasing modes when away from the sample edges. To the best of our knowledge, this is first demonstration of individual access to localized modes of a disorder system despite modal overlap and mode competition. The impact of nonlinearities on localization is a long-standing open question, which we address here in a clear and controlled way. We show unequivocally that the spatial profile of the lasing mode is robust to nonlinear gain saturation, cross-mode saturation and mode competition.
Our method can be easily extended to study localized or extended modes of a 2D disorder systems and 2D-random lasers, as well as 2D-microcavity lasers. The versatility of the method and the degree of control it offers open the way to the exploration of on-ship light transport in scattering media, in the presence of fully-monitored non-uniform gain and loss. In this study, we have intentionally chosen a strongly scattering random laser to operate in the localized regime. Yet the depth of the grooves can be adjusted to further explore different regime of transport. Non-Hermitian optical systems can be investigated as well as light transport control by gain/loss distribution.

## Acknowledgments

We are grateful to Dr. Yossi Abulafia for his help in the fabrication process and the Bar-Ilan Institute of Nanotechnology & Advanced Materials for providing with fabrication facilities. This research was supported by The Israel Science Foundation (Grants No. 1871/15, 2074/15 and 2630/20) and the United States-Israel Binational Science Foundation NSF/BSF (GrantNo. 2015694). B. K. thanks the PBC Post-Doctoral Fellowship Program from the Israeli Council for Higher Education. P. S. is thankful to the CNRS support under grant PICS-ALAMO.

## References:

[1] Özdemir, S. K., Rotter, S., Nori, F. & Yang, L., "Parity-time symmetry and exceptional points in photonics", Nat. Mater. **18**, 783–798 (2019).

[2,] M. A. Miri, A. Alù, "Exceptional points in optics and photonics", Science **363**, eaar7709 (2019)

[3] C. E. Rüter, K. G. Makris, R. El-Ganainy, D. N. Christodoulides, M. Segev, D. Kip, "Observation of parity-time symmetry in optics", Nat. Phys. **6**, 192–195 (2010).

[4] P. W. Anderson, "Absence of diffusion in certain random lattices," Phys. Rev. **109**, 1492–1505 (1958).

[5]X. Wu, J. Andreasen, H. Cao, and A. Yamilov, "Effect of local pumping on random laser modes in one dimension," J. Opt. Soc. Am. B **24**, A26-A33 (2007).

[6] Andreasen, J., Sebbah, P. & Vanneste, C, "Nonlinear effects in random lasers", *J. Opt. Soci Am. B* **28**, 2947–2955 (2011).

[7] Cao, H. et al., "Random laser action in semiconductor powder", Phys. Rev. Lett. **82**, 2278–2281 (1999).

[8] C. Vanneste, P. Sebbah, "Selective Excitation of Localized Modes in active random media" Phys. Rev. Lett. **87**, 183903 (2001).

[9] C. Vanneste, P. Sebbah, and H. Cao, "Lasing with resonant feedback in the diffusive regime", Phys. Rev. Lett. **98**, 143902 (2007).

[10]     J. Andreasen, A. A. Asatryan, L. C. Botten, M. A. Byrne, H. Cao, L. Ge, L. Labonté, P. Sebbah, A. D. Stone, H. E. Türeci, and C. Vanneste, "Modes of Random Lasers", Advances in Optics and Photonics, **3** Issue 1, 88-127 (2011).

[11] Ge, L. et al., "Unconventional modes in lasers with spatially varying gain and loss", Phys. Rev. A **84**, 023820 (2011).

[12] J. Andreasen, C. Vanneste, L. Ge, and H. Cao, "Effects of spatially nonuniform gain on lasing modes in weakly scattering random systems," Phys. Rev. A **81**, 043818 (2010).

[13] H. E. Türeci, L. Ge, S. Rotter, A. D. Stone, "Strong Interactions in Multimode Random Lasers", Science **320**, 643–646 (2008).

[14] Milner, V. & Genack, A. Z., "Photon localization laser: Low-threshold lasing in a random amplifying layered medium via wave localization", *Phys. Rev. Lett.* **94**, 073901 (2005)

[15] X. Wu, J. Andreasen, H. Cao, and A. Yamilov, "Effect of local pumping on random laser modes in one dimension"  J. Opt. Soc. Am. B,  **24** A26 (2007).

[16]  van der Molen, K. L., Tjerkstra, R. W., Mosk, A. P. & Lagendijk, A. Spatial extent of random laser modes. Phys. Rev. Lett. **98**, 143901 (2007).

[17] Tulek, A., Polson, R. & Vardeny, Z. Naturally occurring resonators in random lasing of $\pi$-conjugated polymer films. *Nature Phys* **6,** 303–310 (2010).

[18] X. Jiang and C. M. Soukoulis, "Localized random lasing modes and a path for observing localization", Phys. Rev. E **65**, 025601 (2001).

[19] J. Liu, P. D. Garcia, S. Ek, N. Gregersen, T. Suhr M. Schubert, J. Mørk, S. Stobbe P. Lodahl, "Random nanolasing in the Anderson localized regime", Nature Nanotechnology **9**, 285-289 (2014).

[20] Segev, M., Silberberg, Y. & Christodoulides, D. N., "Anderson localization of light", *Nat. Photon.* **7**, 197–204 (2013).

[21]X. Wu, J. Andreasen, H. Cao, and A. Yamilov, "Effect of local pumping on random laser modes in one dimension," J. Opt. Soc. Am. B **24**, A26-A33 (2007).

[22]  N. Bachelard, J. Andreasen, S. Gigan, P. Sebbah, "Taming random lasers through active spatial control of the pump", Phys. Rev. Lett., **109**, 033903 (2012).

[23] N. Bachelard, S. Gigan, X. Noblin, P. Sebbah, "Adaptive pumping for spectral control of random lasers", Nature Physics **10**, 426–431 (2014).

[24] S. F. Liew, B. Redding, L. Ge, G. S. Solomon, and H. Cao, "Active control of emission directionality of semiconductor microdisk lasers", Appl. Phys. Lett. **104**, 231108 (2014).

[25] F. Gu, F. Xie, X. Lin, S. Linghu, W. Fang, H. Zeng, L. Tong , S. Zhuang. "Single whispering-gallery mode lasing in polymer bottle micro resonators via spatial pump engineering" Light Science and Applications **6**, 17061 (2017).

[26] Y. Qiao, Y. Peng, Y. Zheng, F. Ye and X. Chen, "Adaptive pumping for spectral control of broadband second-harmonic generation", Opt. Lett., **43**, 787 (2018).

[27] L. Ge, O. Malik, H. E. Türeci, "Enhancement of laser power-efficiency by control of spatial hole burning interactions" Nat. Photonics . **8,** 871 (2014).

[28] T. Hisch, M. Liertzer, D. Pogany, F. Mintert, and S. Rotter, "Pump-controlled directional light emission from random lasers", Phys. Rev. Lett. , **111**, 023902 (2013).

[29] J.-H. Choi, S. Chang, K.-H. Kim, W. Choi, S.-J. Lee, J. M. Lee, M.-S. Hwang, J. Kim, S. Jeong, M.-K. Seo, W. Choi, and H.-G. Park, "Selective Pump Focusing on Individual Laser Modes in Microcavities", ACS Photonics **5**, 2791 (2018).

[30] A. Cerjan, B. Redding, L. Ge, S. F. Liew, H. Cao, and A. D. Stone, "Controlling mode competition by tailoring the spatial pump distribution in a laser: a resonance-based approach" Opt. Express . **24**, 26006 (2016).

[31] L. Ge, "Selective excitation of lasing modes by controlling modal interactions", Opt. Express **23** 30049 (2015).

[32] A. Cerjan, B. Redding, L. Ge, S. F. Liew, H. Cao, and A. D. Stone, "Controlling mode competition by tailoring the spatial pump distribution in a laser: a resonance-based approach", Opt. Express **24**, 26006 (2016).

[33] M. Lebental, N. Djellali, C. Arnaud, J.-S. Lauret, J. Zyss, R. Dubertrand, C. Schmit, and E. Bogomolny, "Inferring periodic orbits from spectra of simply shaped microlasers", Phy. Rev. **A76**, 023830 (2007).

[34] Riboli F, Caselli N, Vignolini S, Intonti F, Vynck K et al. Engineering of light confinement in strongly scattering disordered media. Nat Mater. **13**: 720–725,(2014).

[35] Edwards, J. T., Thouless, D. J., "Numerical studies of localization in disordered systems", Journal of Physics C: Solid State Physics,  **5**, Issue 8, 807-820 (1972).

[36] Türeci H E, Stone A D and Collier B , Self-consistent multi-mode lasing theory for complex or random lasing media Phys. Rev. **A 74,** 043822 (2006).

[37] Fox, A.G., Li, T,  "Resonant modes in a maser interferometer". *Bell Syst. Tech. J*. **40** (2): 453–488,(1961).

[38] P Sebbah, C Vanneste, "Random laser in the localized regime", Physical Review B **66** (14), 144202 (2002).

[39] Andreasen, J. *et al*. "Partially pumped random lasers", *Int. J. Mod. Phys. B* **28**, 1430001 (2014).

[40] Ge, L., Liu, D., Cerjan, A., Rotter, S., Cao, H., Johnson, S. G., Türeci, H. E., Stone, A. D., "Interaction-induced mode switching in steady-state microlasers", Opt. Express, **24**, 41−54 (2006).

# Methods:

## Device Fabrication:

Our 1D photonic disorder structure consists of randomly distributed subwavelength air grooves on a PMMA-DCM layer as shown by schematics in figure1(a). They are fabricated on a 600 nm layer of poly (methyl methacrylate) (PMMA) passive host medium doped with 5 % weight of DCM (4-dicyanomethylene-2-methyl-6-(4-dimethylaminostyryl- 4H-pyran) laser dye, whose fluorescence spectrum is centered around 600 nm. Because of good fluorescence quantum yield and large stoke shift (100 nm) it did not reabsorb the emitted light. We use PMMA with a molecular weight of 49500 at a concentration of 6 % weight in anisole. A layer of PMMA-DCM dye-doped polymer is obtained by spin-coating the doped-polymer solution at 1000 rpm for 60 sec and post baked at 120 ºC for 2 hours in oven on a fused silica transparent glass wafer (Edmund optics) thickness 1 mm and having refractive index of 1.45 compare to 1.65 for PMMA-DCM layer [1]. Refractive index of the PMMA-DCM layer is higher than fused silica substrate and air, this ensure wave guiding of the fundamental optical mode around 600 nm. Disordered structures are fabricated by electron-beam lithography technique (CRESTEC/CABL-9000C). Electron beam exposure of bare PMMA-DCM layer generate significant charging effects, as expected in case of insulating substrate. To avoid that, we add a 20 nm layer of conductive polymer (ESpacer) to eliminate the charging effect. After E-beam writing conductive polymer was removed with DI (Deionized) water by immersion for 40 s. To design 1D disorder scattering system, parallel grooves, each 200 nm-wide and 50 μm-long, are carved randomly on PMMA-DCM layer using E-beam lithography, which gives dielectric layers of average length 8 μm separated by an air gap of 200 nm as shown by optical microscope image taken at 10x in figure 1(c). We can vary the width and depth of air grooves by changing current and exposure time of electron beam. It give us flexibility to change scattering strength without changing disorder configuration. High-resolution SEM images if sample are shown in supplementary information (S1).

**Experiment Set-up**: Schematics of experimental setup shown in figure 2(f). Experiment setup consist of an incident light from frequency-doubled ND:YAG laser (EXPLA PL2230: 532 nm, 20 ps, maximum output energy 28 mJ, repetition rate 10 Hz) is expanded 5x to be spatially modulated by a 1952×1088 pixels reflective SLM (Holoeye HES 6001, pixel 8.0 µm). The SLM is placed in the object plane of telescope with 5x reduction and is imaged on the sample. This set up provide a 500 µm long and 50 µm wide laser strip line nearly diffraction free modulation down to large rectangular pixels. Disorder structure is precisely aligned with the laser strip line under a fixed stage Zeiss (AxioExaminer A1) microscope and imaged using Andor zyla sCMOS (22 mm diagonal view, 6.5 m pixel size) CCD camera microscope. The laser emission is collected via an optical fiber connected to Horiba iHR550 imaging spectrometer equipped with a 2400 $mm^{-1}$ grating and samples are images by 10X objective on a synapse CCD detection system (sampling rate 1MHz, 1024 X 256 pixels, 26 m pixel pitch). The entrance slit is 50 µm. The resulting spectral resolution is 20 pm. The integration time is 1 sec.

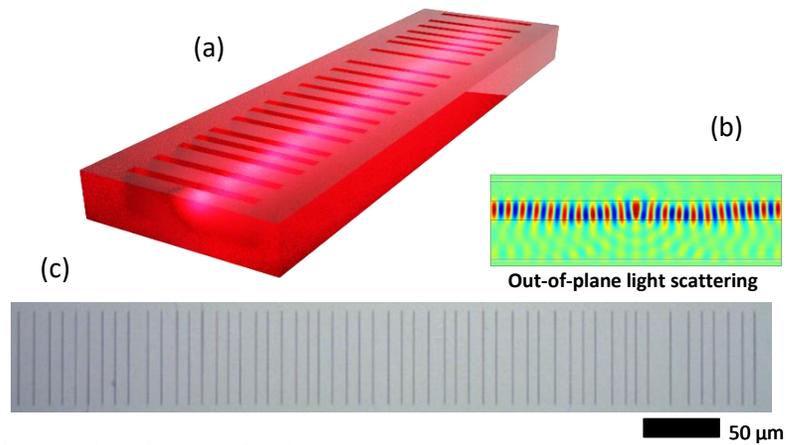

***Figure 1*: The "plastic random laser".**
The polymer-based solid-state random laser is a (DCM) dye-doped 600 nm-thick layer of polymer (PMMA) spin-coated on fused silica. The scattering is provided by 125 parallel grooves (600 nm-deep, 120 nm-wide, 50 µm long) randomly distributed along the 1 mm-long sample with an average period of 8 µm. (a) Artist view of the random laser, which is optically pumped from below by a laser strip line. Laser oscillations result from multiple scattering on the grooves and amplification between the grooves. Emission occurs essentially along the waveguide direction, while a small percentage is scattered by the grooves and emitted transversally. Out-of-plane scattering at the groove position allows near-field imaging within the random laser. (b) 2D-simulation of first optical mode propagating within the waveguide and scattered at a groove. Bottom layer: Fused silica substrate (refractive index, $n$=1.45); Middle layer: Dye-doped polymer ($n$=1.65); top layer: air ($n$=1). (c) 20x microscope image of the sample.

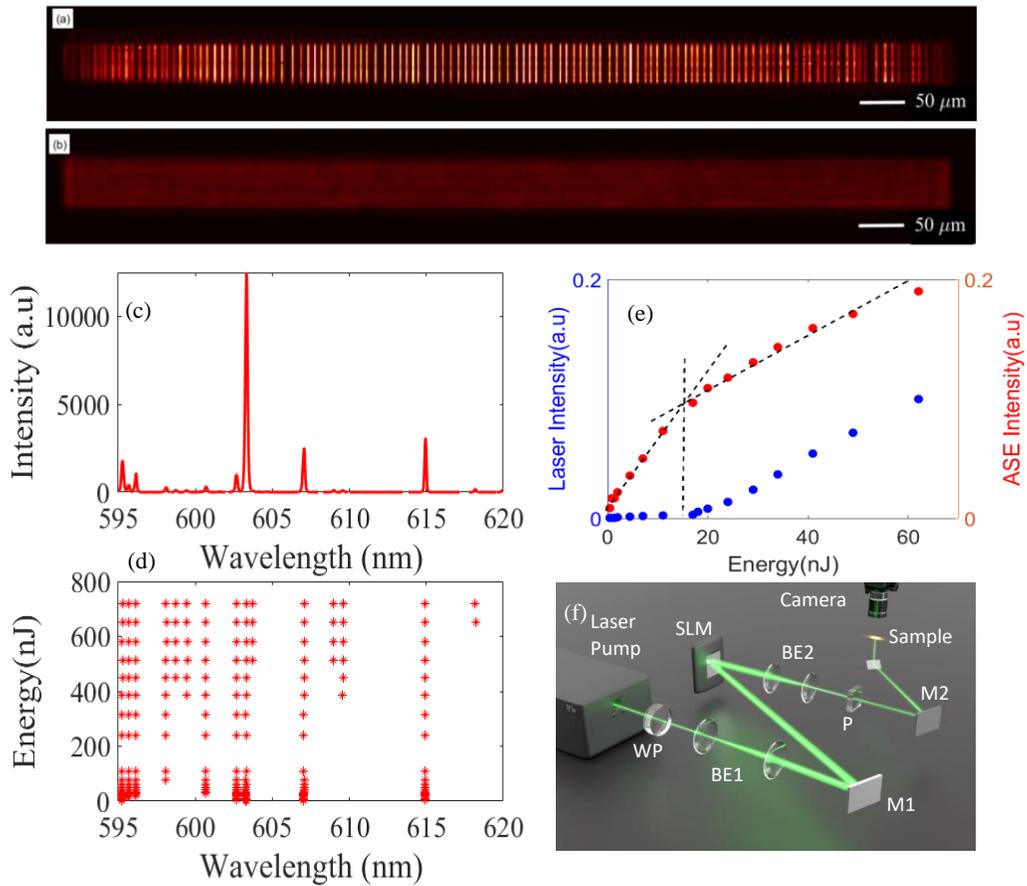

**Figure 2: Laser characteristics under uniform pumping.**
Near-field optical microscope image of random laser out-of-plane emission when (a) above threshold and (b) pumped below threshold. (c) Multimode laser emission spectrum under uniform pumping above threshold at pump energy 700 nJ, averaged over 50 shots. (d) Spectral position of lasing modes plotted against pump energy. (d) Integrated output intensity (red stars) measured vs excitation power and integrated amplified spontaneous emission (ASE) (blue stars) vs excitation power. (e) Measure of amplified spontaneous emission (ASE) and laser intensity vs. pump energy. Lasing onset and ASE clamping are observed at threshold, for a pump energy 18nJ. (f) **Schematics of experiment setup:** The
28 ps pulse of a doubled Diode Pumped Solid State mode-locked Nd:YAG laser (Ekspla PL2231-50) is first expanded on a computer-driven LCOS amplitude spatial light modulator (SLM) (Holoeye HES 6001). Input laser polarization is controlled by half-wave plate (WP). The modulated intensity pattern is reduced and imaged on the back of the sample. A fixed-stage optical microscope (Zeiss Axio-examiner) (not represented) magnifies (10x) the near-field at the surface of the sample onto a scientific CMOS camera (Andor Zyla 4.2+). M1 and M2: Mirrors. BE1 and BE2: Beam expanders. P: Polarizer.

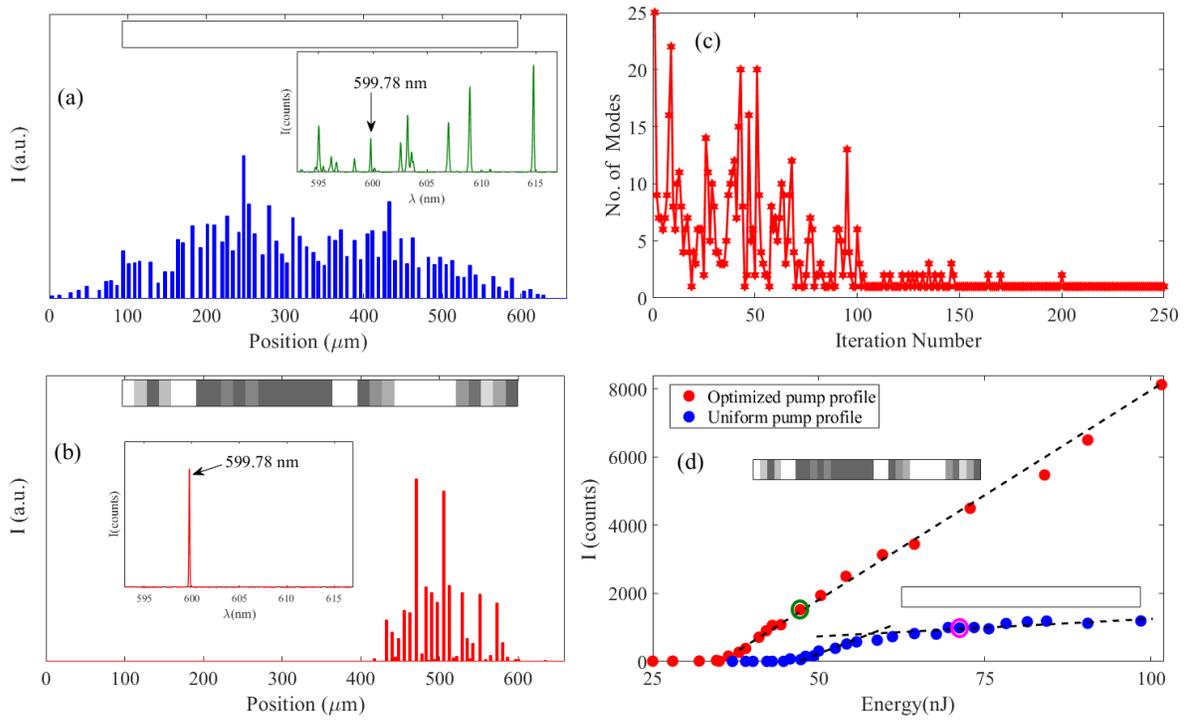

**Figure 3: Lasing mode under uniform pump profile vs. optimized pump profile.**
Left panels: Spatial field-intensity distribution within the random laser (each bar integrates light intensity scattered from one air groove), corresponding emission spectrum (inset) and pump intensity profile (top inset). (a) Under uniform pumping (multimode operation) (pump energy is 71.2 nJ); (b) Under optimized pump profile for singlemode laser emission at 599.78 nm (pump energy reduces to 41.2 nJ). (c) Number of lasing modes vs. iteration number during optimization process at 599.78 nm. (d) Laser characteristic for the lasing mode @599.78 nm, (blue) in multimode operation, (red) in singlemode operation when optimized pump profile is applied. In multimode regime, the threshold of the mode @599.78 nm is 47.3 nJ and its slope efficiency is 20 (counts/nJ). In singlemode regime, its threshold is reduced to 35 nJ, and its slope efficiency increases to 123 (counts/nJ). The purple circle (resp. green circle) indicates the pump energy level at which optimization is started (resp. is finished). Note the saturation of the lasing mode resulting from mode competition in multimode regime (blue curve), which disappears when the mode is optimally pumped.

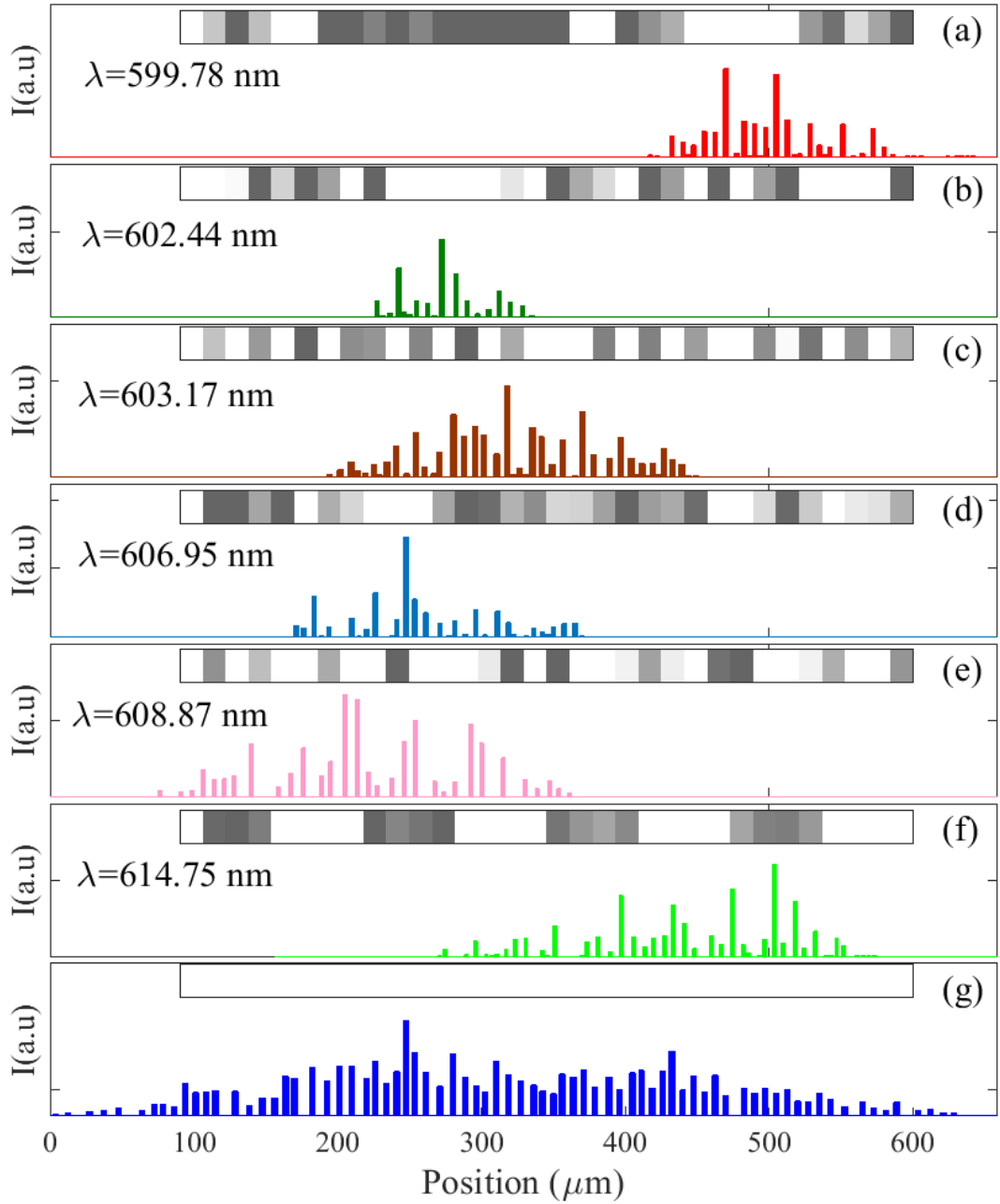

**Figure 4: Selective pumping of localized lasing modes.**
(a-f) Spatial distribution of laser intensity within the random structure for singlemode operation of the random laser at λ =599.78 nm, 602.44 nm, 603.17nm, 606.95 nm, 608.87 nm and 614.75 nm. Singlemode operation has been achieved by iterative optimization of the pump profile. The corresponding optimized pump profile is reproduced in greyscale on top of each figure (white: maximum intensity; Black: no pumping). The spectral side lobe rejection obtain with this method is 63.18 dB, 64.49 dB, 60.56 dB, 58.72 dB, 66.41 dB and 69.87 dB, respectively. (g) Spatial distribution of laser intensity for uniform pumping and multimode operation.

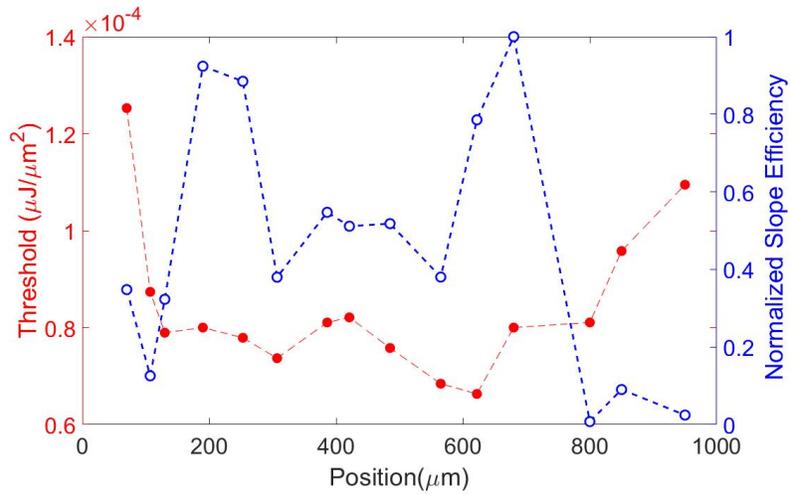

**Figure 5:** Lasing threshold (red dots) and slope efficiency (blue open circles) vs. mode barycenter position along the sample, of individually-excited lasing modes (from left to right) @λ= 622.12 nm, @609.02 nm, @605.05 nm, @606.91 nm, @595.48 nm, @604.73 nm, @606.49 nm, @619.04 nm, @614.48 nm, @612.60 nm, @617.15 nm, @606.35 nm, @598.66 nm, @602.74 nm is plotted. All lasing modes have been pumped locally over 80 μm.

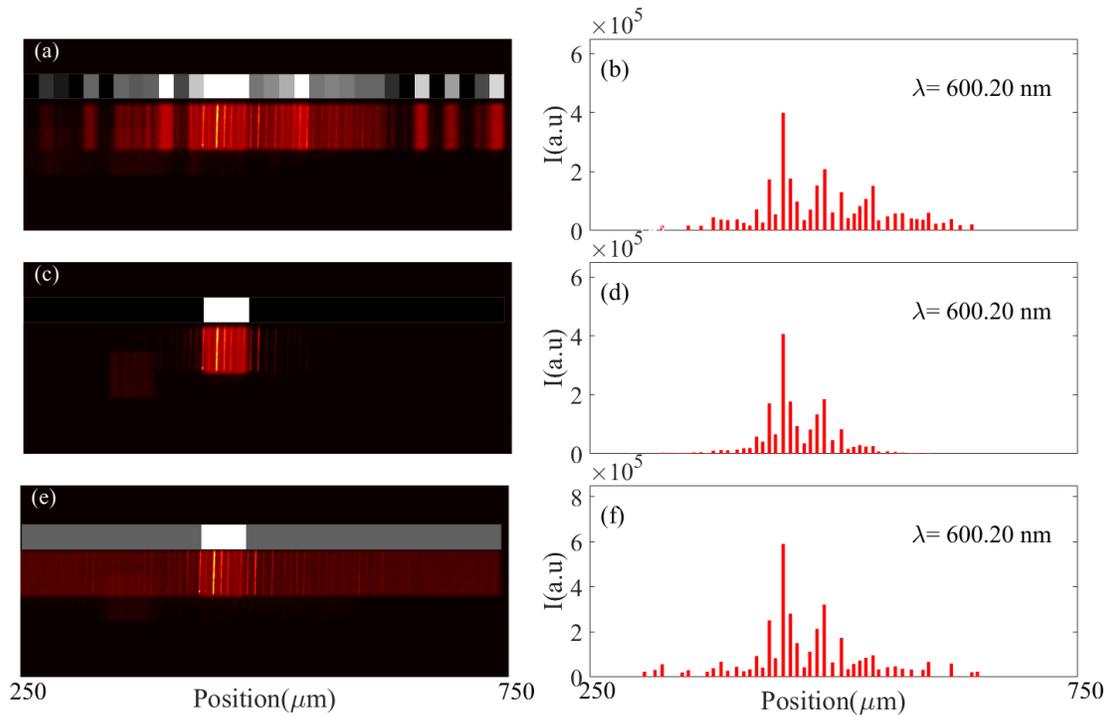

**Figure 6: Optimized vs. local pumping of a lasing mode localized near the MIDDLE of the sample.**
Left panels: Near-field optical microscope image (colorscale) and pump profile (greyscale) for mode selected @600.20 nm, **which is localized near the middle of the sample**. Right panels: corresponding spatial distribution of random laser light intensity within the random laser (fluorescence has been subtracted). (a)(b) Optimally-pumped laser mode @600.20 nm (pump energy is 102 nJ); (c)(d) Locally-pumped mode @600.20 nm (pump length is 46 µm, pump energy is 75 nJ. (e)(f) Same as (c)(d) after losses have been compensated by adding a small uniform pump. Correlation coefficient between (b) and (d) distributions is 82% and rises to 96% between (b) and (f).

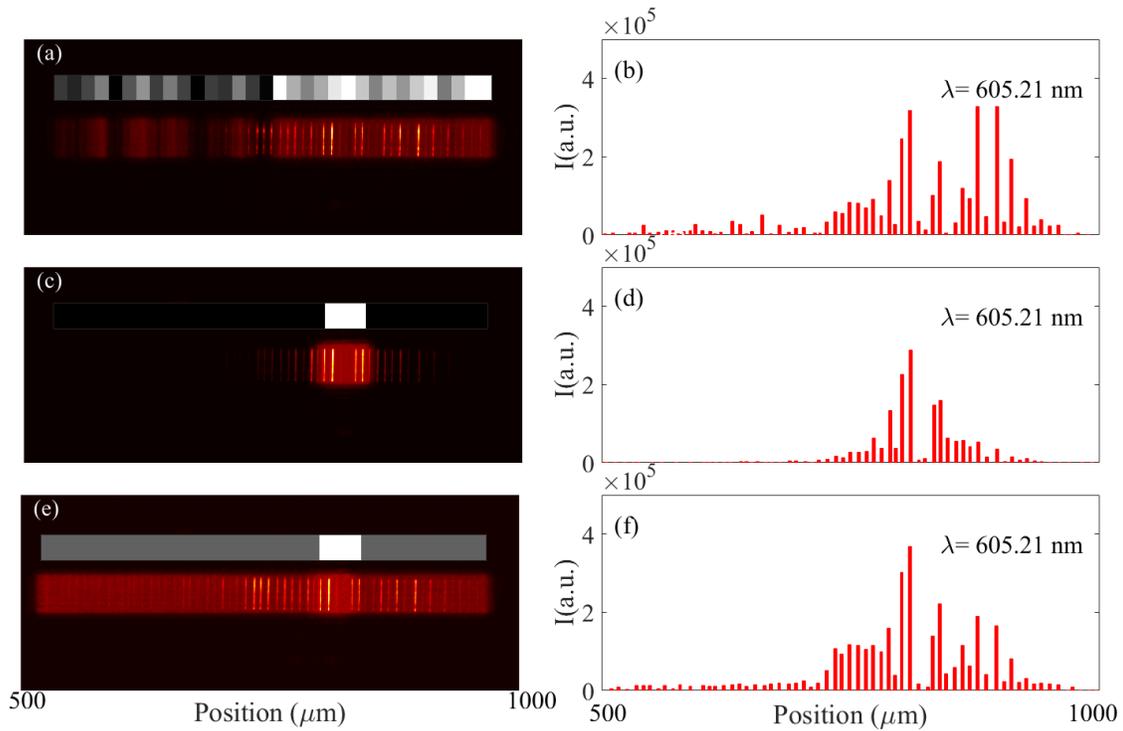

**Figure 7: Optimized vs. local pumping of a lasing mode localized near the EDGE of the sample.**
Left panels: Near-field optical microscope image (colorscale) and pump profile (greyscale) for mode selected @605.21 nm, **which is localized near the middle of the sample**. Right panels: corresponding spatial distribution of random laser light intensity within the random laser (fluorescence has been subtracted). (a)(b) Optimally-pumped profile @605.21nm (140 nJ); (c)(d) Locally pumped mode @605.21 nm (pump length is 42 µm) (singlemode @605.21 nm, 142 nJ). (e)(f) Same as (c)(d) after losses have been compensated by adding a small uniform pump. Correlation coefficient between (b) and (d) distributions is 56%, while it rises to 90% between (b) and (f).

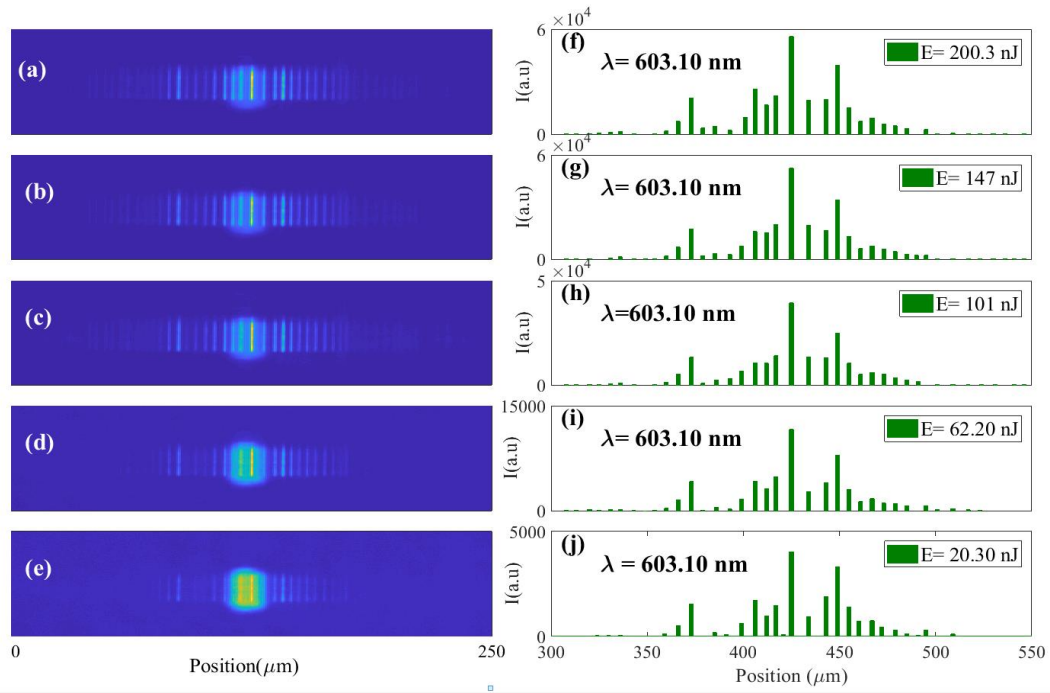

**Figure 8: Localized lasing modes in presence of gain nonlinearity.**
After optimization, length of 32 pixels optimal pump profile is minimized by removing pixels from both sides until we achieve minimum pump length sufficient to individually excite target mode λ= 603.10 nm a higher pump energy. (a-e) Shows optical microscope image (10X) of the spatial field distribution of lasing mode at λ = 603.10 nm excited by a pump length L= 10 μm at pump power of 42.7, 83.5, 130, 165.3, 200.3 nJ (bottom to top). (f-j) Intensity profile (bar plots) of the corresponding left image.

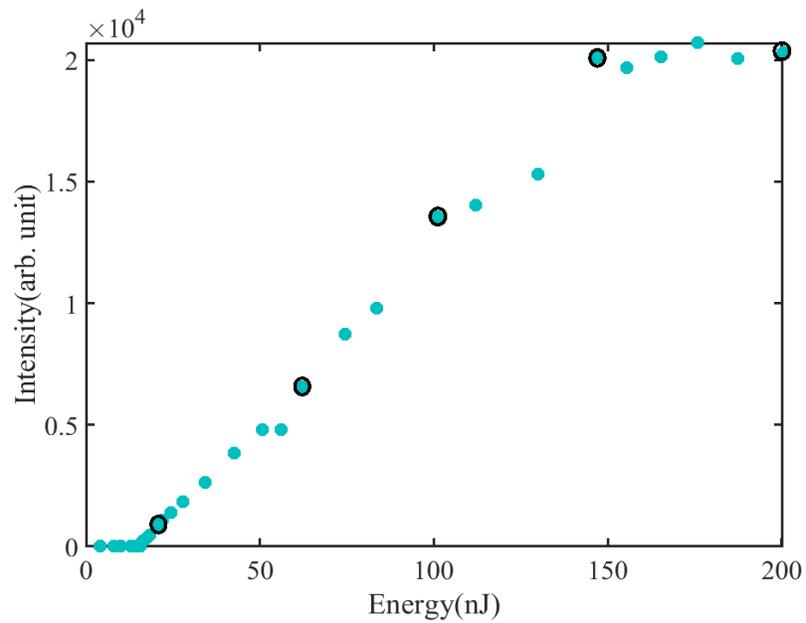

**Figure 9: Gain saturation in singlemode operation.**
Lasing characteristic for optimally-selected lasing mode @603.10 nm, operating in singlemode regime. Threshold is 14 nJ. Gain saturation is reached at 147 nJ. Black circles indicate the pump energies at which the mode intensity distribution is shown in Fig. 8.

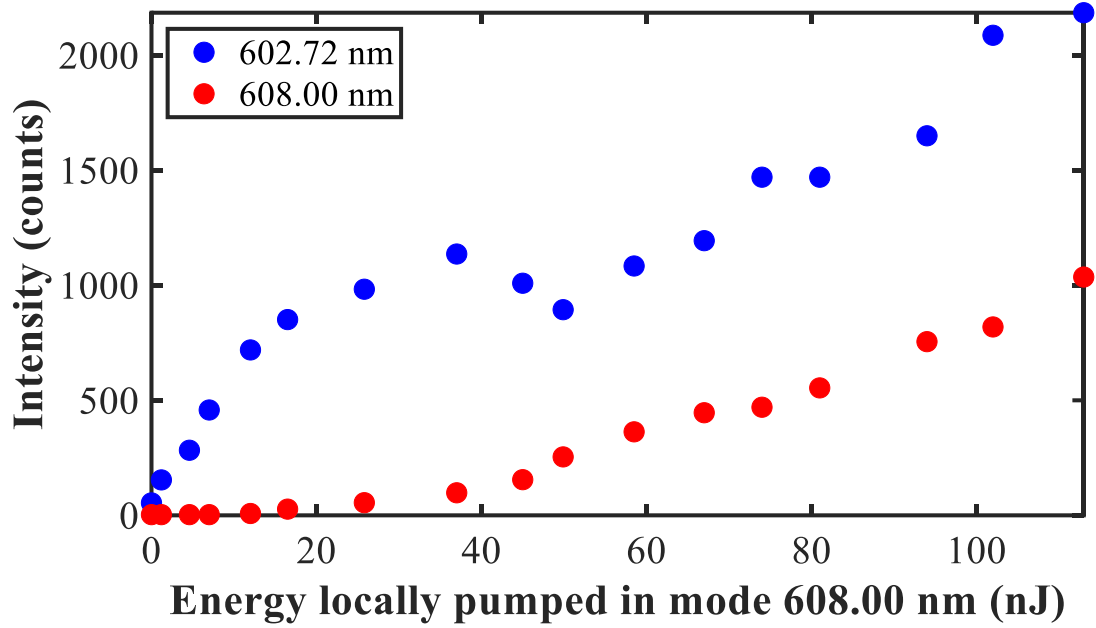

**Figure 10:** Nonlinear competition and cross saturation effect between individually selected lasing mode1(608.00 nm) and mode2(602.72 nm) with corresponding pump profile P1 and P2 : Evolution in peak intensity of mode 1 and mode 2 with progressive increase in pump energy of profile P1 at a constant above threshold pump energy (48 nJ) in profile P2.

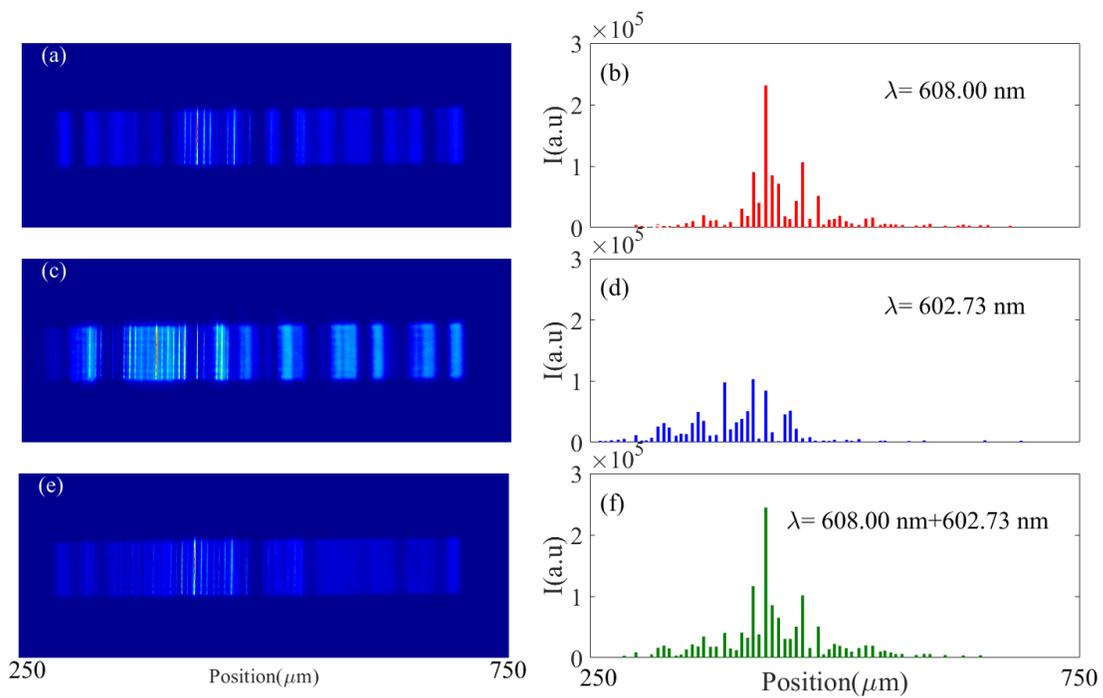

**Figure 11: Mode competition.**
(a,b) Near field optical microscope image and intensity profile of individually selected mode 608.00 nm. (c,d) Near field optical microscope image and intensity profile of individually selected mode 608.00 nm. (e,f) Near field optical microscope image and intensity profile recorded when both mode 608.00 nm and 602.72 nm lasing together and competing for the gain

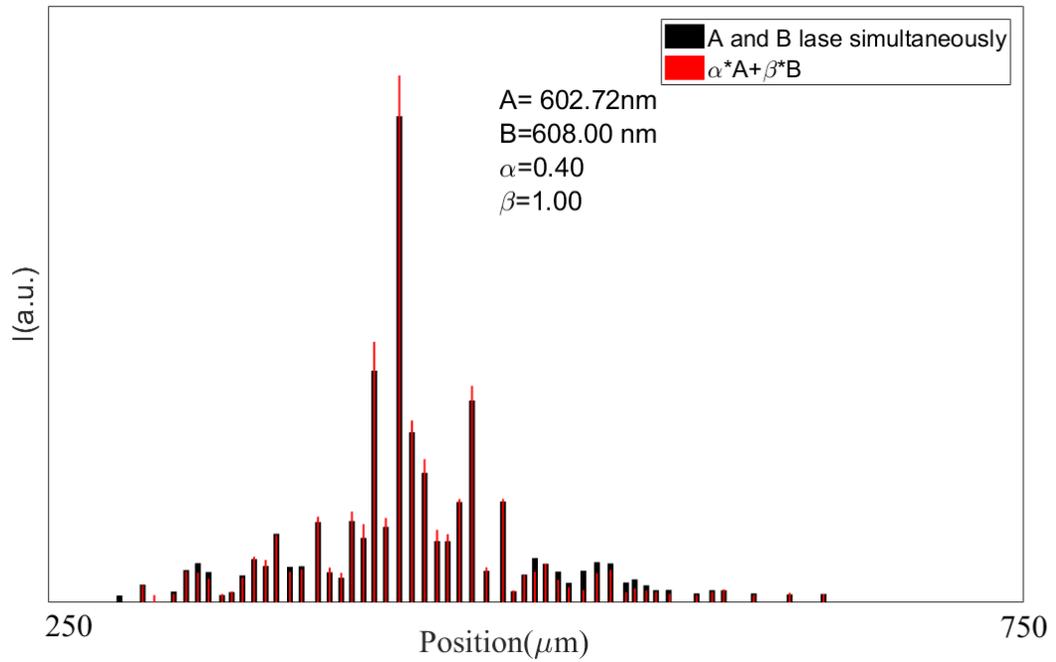

**Figure 12 :** Comparison of the sum of the intensity profile of individually selected mode 1 and mode 2 (red bar) with intensity profile obtained when both modes lasing together under the effect of cross saturation and gain competition (black bar).